\documentclass[letterpaper,journal]{IEEEtran}

\usepackage{amsmath,amsthm,amssymb}
\usepackage{cite}
\usepackage{microtype}

\usepackage{bm}
\usepackage{bbm}
\usepackage{algorithm, algorithmic}
\usepackage{array}
\usepackage{textcomp}
\usepackage{stfloats}
\usepackage{url}
\usepackage{verbatim}
\usepackage{xcolor}
\def\BibTeX{{\rm B\kern-.05em{\sc i\kern-.025em b}\kern-.08em
    T\kern-.1667em\lower.7ex\hbox{E}\kern-.125emX}}
\usepackage{balance}

\usepackage{graphicx}
\usepackage{epsfig}
\usepackage{diagbox}

\ifCLASSOPTIONcompsoc
  \usepackage[caption=false,font=normalsize,labelfont=sf,textfont=sf]{subfig}
\else
  \usepackage[caption=false,font=footnotesize]{subfig}
\fi

\usepackage{lipsum}
\usepackage{mathtools}
\usepackage{cuted}

\DeclareMathOperator*{\argmax}{arg\,max}

\begin{document}
\title{Pearcey-Inspired Quartic Wavefront Shaping for Obstructed Near-Field Multi-User Communications}

\author{Mingcan Qin,
        Jing Chen,
        and Yifeng Qin,~\IEEEmembership{Member,~IEEE}

\thanks{Corresponding author: Yifeng Qin.}
\thanks{M. Qin is with Suzhou City University, Suzhou 215200, Jiangsu, China (e-mail: 769132289@qq.com).}
\thanks{J. Chen and Y. Qin are with the Peng Cheng Laboratory, Shenzhen, 518052, China (e-mails: chenj12@pcl.ac.cn, ee06b147@gmail.com).}

}

\maketitle

\begin{abstract}
Radiative near-field (RNF) beamforming is vulnerable to Fresnel-zone blockages.
This letter proposes an obstruction-unaware wavefront shaping strategy inspired by catastrophe optics: a calibrated quartic phase generates a Pearcey-inspired wavepacket that improves effective multi-user channel separability in the tested partially blocked geometries.
The quartic beam is calibrated in free space without obstruction information, while digital zero-forcing (ZF) uses the resulting effective channel to evaluate common SINR under a shared total transmit-power constraint.
Numerical results demonstrate a peak sampled common-SINR gain of $4.29$~dB over conventional focusing when closely spaced users experience strong central blockage.
\end{abstract}


\section{Introduction}
\IEEEPARstart{T}{HE} evolution toward 6G networks has spurred significant interest in extremely large-scale antenna arrays (ELAA) operating in the radiative near-field (RNF) region~\cite{ref_nf_mag, You2021}.
Unlike the planar wavefronts assumed in far-field communications, the spherical wavefronts in the RNF region enable beam focusing at specific locations, unlocking the range dimension for spatial multiplexing~\cite{ref_nf_tutorial}.
This capability allows a base station (BS) to serve multiple users located at the same angle but different distances, effectively mitigating multi-user interference (MUI) through range-division multiple access~\cite{ref_elaa_access}.

However, the theoretical gains of RNF focusing are highly sensitive to the propagation environment.
In practical scenarios, line-of-sight (LoS) paths are frequently partially blocked by finite obstructions (e.g., lamp posts, human bodies, or vehicles).
While complete blockage is detrimental, partial blockage---diffraction by finite obstacles---induces amplitude fluctuations and phase distortions~\cite{Guerboukha2024Curving}.
In multi-user (MU) systems, these distortions can reduce the separability of users' channel vectors.
When users are closely spaced relative to the array's depth-of-focus scale, blockage can weaken the effective channel's smallest singular mode.
Consequently, zero-forcing (ZF) requires a larger power expenditure to separate users, reducing their common SINR under a fixed transmit-power budget~\cite{ref_xla_mag}.

Existing solutions to enhance robustness include channel-assisted beam and blockage prediction~\cite{ref_nf_model} or deploying reconfigurable intelligent surfaces (RIS) to create additional propagation paths around blockages~\cite{ref_block_1,ref_Yifeng_IOS}.
Alternatively, diffraction-free beams, such as Bessel~\cite{Jornet2023Bessel,Qin2025Roadmap,Bodet2024Measurements} and Airy beams~\cite{Guerboukha2024Curving,Darsena2025Fundamentals,LiuConvex,Zhao2025DataCenter}, have been explored for their self-healing properties.
Bessel beams distribute energy along an extended propagation axis, making them useful for coverage; serving co-angular users at distinct ranges also requires sufficiently different channel responses.
Airy beams, owing to their auto-bending trajectories, are particularly suited to half-space blockages such as walls or knife-edges, where the main lobe can be steered around the geometric shadow~\cite{Darsena2025Fundamentals,Qin2026AiryHalfspace}. For the symmetric, on-axis finite obstructions considered here, an on-axis-centered wavepacket with even symmetry in $x$ allows contributions from both sides of the obstruction to reach the user axis.
In MU systems, obstruction robustness depends on both the received power and the distinct spatial signatures of the users. We therefore examine the joint effective channel under blockage, with the aperture weights designed without obstruction information.

In this letter, we propose a Pearcey-inspired quartic wavefront shaping strategy for RNF MU communications.
Drawing from catastrophe optics~\cite{Ring2012Auto}, we exploit the Pearcey integral, which inherently contains a quartic phase term.
Pearcey-type fields exhibit auto-focusing and self-healing behavior in optical experiments~\cite{Ring2012Auto}. These properties motivate using a quartic phase to redistribute the field around the propagation axis, so that reception beyond a finite obstruction can draw on contributions passing through its open sides.

We demonstrate that superimposing a calibrated quartic phase onto the conventional focusing phase can improve the weakest mode of the blocked multi-user channel and reduce the transmit power required for ZF interference cancellation.
The aperture design is obstruction-unaware; digital ZF uses the post-blockage effective channel.

Our contributions are as follows:
\begin{enumerate}
    \item We establish an obstruction-unaware fair comparison protocol between conventional near-field focusing and Pearcey-inspired quartic shaping, where beam calibration is performed in free space and both transmitted signals satisfy the same total array-power constraint.
    \item We use two dimensionless parameters, $W_\text{obs}/r_F$ and $\mu = \Delta z/\Delta z_\text{DOF}$, to map the ZF common-SINR advantage for the considered geometry.
    \item We relate the SINR gain to the weakest singular mode of the power-normalized effective channel, showing how improved multi-user separability can compensate for the intrinsic focal-intensity penalty of the quartic phase under blockage.
\end{enumerate}

\begin{figure}[!t]
    \centering
    \includegraphics[width=1\linewidth]{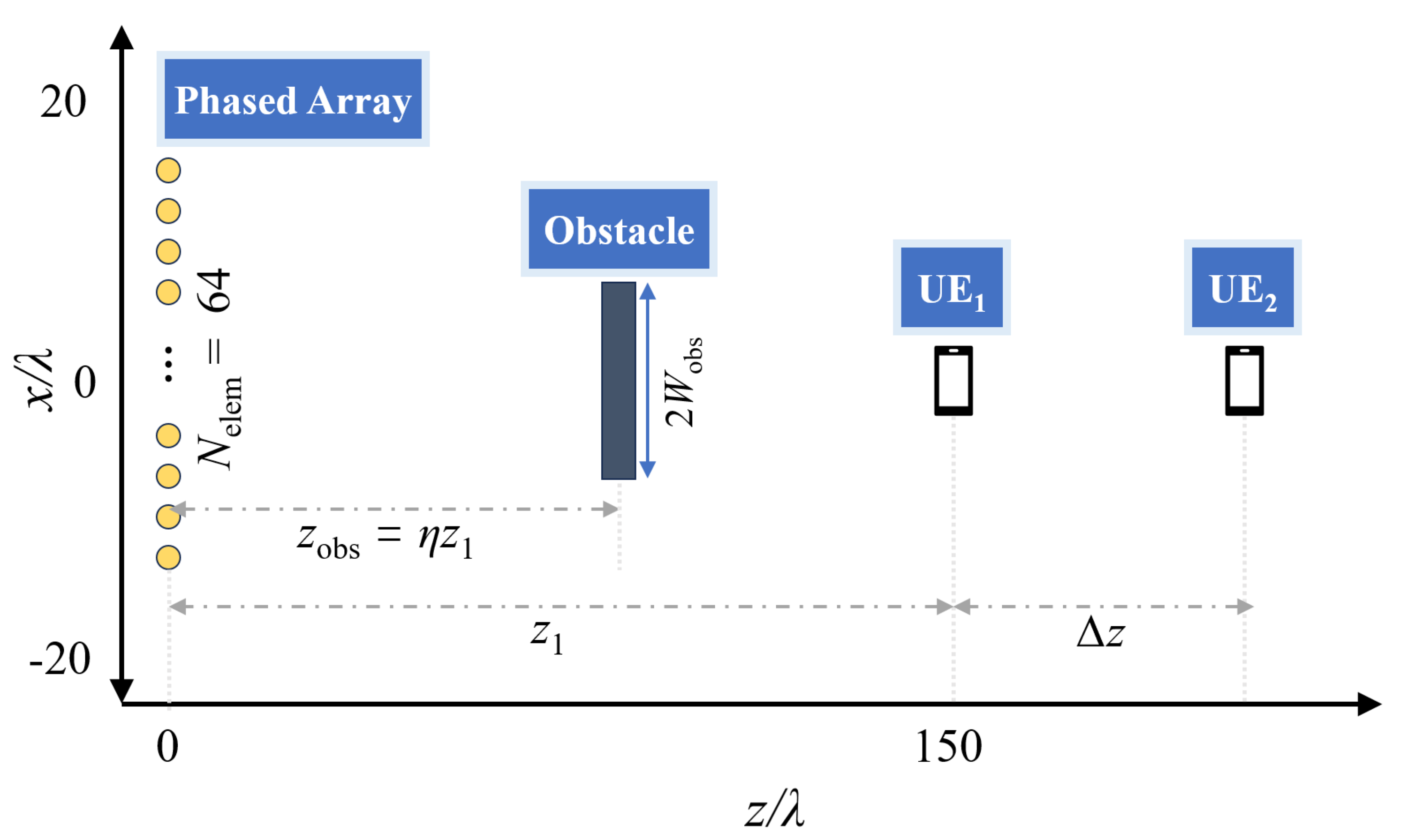}
    \caption{System geometry illustrating the ULA with aperture $D = (N_{\text{elem}}-1)\,d$, the two co-angular users $\text{UE}_1$ and $\text{UE}_2$ separated by $\Delta z$, and the finite obstruction of width $2W_{\text{obs}}$ located at $z_{\text{obs}}$.}
    \label{fig:system_model}
\end{figure}

\section{System Model and Propagation Analysis}
\label{sec:system_model}

\subsection{System Geometry and Physical Scaling}
\label{subsec:geometry}

Consider a radiative near-field (RNF) downlink scenario in a two-dimensional $xz$-plane.
A Base Station (BS) is equipped with a Uniform Linear Array (ULA) located on the $x$-axis at $z=0$.
The array consists of $N_{\text{elem}} = 64$ elements with inter-element spacing $d = 0.49\lambda$, spanning a total physical aperture $D = (N_{\text{elem}} - 1)\,d \approx 31\lambda$.

The system serves two single-antenna users, denoted as $\text{UE}_1$ and $\text{UE}_2$, located along the array's normal direction (the $z$-axis).
As illustrated in Fig.~\ref{fig:system_model}, $\text{UE}_1$ is located at $(0, z_1)$, and $\text{UE}_2$ is located at $(0, z_2)$, where $z_2 = z_1 + \Delta z$.
Here, $\Delta z > 0$ denotes the range separation between the users.

A finite obstruction is located at a distance $z_{\text{obs}} = \eta z_1$ ($0 < \eta < 1$) from the BS, parallel to the ULA.
This obstruction is modeled as an opaque strip centered on the propagation axis with a physical half-width $W_{\text{obs}}$ (total width $2W_{\text{obs}}$).

We characterize obstruction size and user separation using two dimensionless parameters for this geometry.

\subsubsection{Fresnel Radius at Obstruction}
The diffraction pattern depends on the obstruction size relative to the propagation scale.
We use the Fresnel transverse reference scale
\begin{equation}
    r_F = \sqrt{\lambda z_{\text{obs}}},
    \label{eq:r_F}
\end{equation}
where $\lambda$ is the carrier wavelength and $r_F$ characterizes propagation from the array to the obstruction plane.
Normalizing $W_{\text{obs}}$ by $r_F$ expresses the obstruction width relative to the transverse diffraction scale. Together with $\eta$, it specifies the obstruction size and position used in the comparison.

\subsubsection{Normalized Range Separability}
The finite aperture $D$ limits range discrimination through its depth of focus~\cite{ref_dof}.
We adopt the following depth-of-focus reference scale:
\begin{equation}
    \Delta z_{\text{DOF}} \approx \frac{2\lambda z_1^2}{D^2}.
    \label{eq:DOF}
\end{equation}
The normalized range separability $\mu$ is then:
\begin{equation}
    \mu = \frac{\Delta z}{\Delta z_{\text{DOF}}}.
    \label{eq:mu}
\end{equation}
For $\mu\lesssim1$, the users are separated by no more than this depth-of-focus scale, so their focal responses can overlap strongly. Increasing $\mu$ gives the array more range separation for distinguishing the users.

\subsection{Propagation and Effective Channel Model}
\label{subsec:propagation}

\subsubsection{Fresnel Propagation}
We use scalar angular-spectrum propagation with the $e^{j\omega t}$ time convention.
Let $E_0(x)$ be the aperture profile obtained by linear interpolation of the discrete array weights, with support $[-D/2,D/2]$.
The propagated field is
\begin{equation}
    E(x,z)=\mathcal{F}^{-1}\!\left\{\mathcal{F}[E_0]e^{-jk_z z}\right\},\; k_z=\sqrt{k_0^2-k_x^2},
    \label{eq:fresnel_prop}
\end{equation}
where $k_0=2\pi/\lambda$ and only propagating components $|k_x|\leq k_0$ are retained. The paraxial limit has the Fresnel kernel $\exp[-jk_0(x-x')^2/(2z)]$, consistent with the conjugate focusing phase below. The numerical operator uses zero padding; interpolation is linear, with power normalized at the discrete array elements.
\subsubsection{Obstruction Modeling}
The finite obstruction at $z_{\text{obs}}$ imposes a multiplicative mask $M(x)$ on the incident field:
\begin{equation}
    M(x) = \begin{cases}
    0, & |x| < W_{\text{obs}} \\
    1, & \text{otherwise}
    \end{cases}.
    \label{eq:mask}
\end{equation}
The propagation is modeled as a two-step process: the field first propagates from the aperture to the obstruction plane, is masked by $M(x)$, and then the surviving field propagates the remaining distance $(z - z_{\text{obs}})$ to the user locations.

\subsection{Effective Channel, Precoding, and Metrics}
\label{subsec:metrics}

\subsubsection{Effective MU Channel}
For $K=2$ users, let $\mathbf w_j\in\mathbb C^{N_\text{elem}}$ be the unit-norm array weight vector for beam $j$, and define $\mathbf W=[\mathbf w_1,\mathbf w_2]$.
The effective coefficient $h_{ij}$ is the complex field at user $i$ produced by beam $j$ through the two-step propagation and obstruction mask:
\begin{equation}
    h_{ij} = E_{\mathbf{w}_j}(0, z_i), \quad \mathbf{H} = \begin{bmatrix} h_{11} & h_{12} \\ h_{21} & h_{22} \end{bmatrix} \in \mathbb{C}^{2 \times 2}.
    \label{eq:effective_channel}
\end{equation}
Here $h_{ij}$ represents the post-beamforming effective scalar channel, obtained by applying the $N_\text{elem}$-dimensional weight vector $\mathbf{w}_j$ to the spatial channel and propagating through the obstruction model.

\subsubsection{ZF Common-SINR}
Zero-forcing (ZF) precoding is adopted to examine how blockage-induced changes in the weakest channel mode affect simultaneous user service. ZF cancels inter-user interference by inverting the effective channel; under a fixed transmit-power budget, a larger inversion cost reduces the signal level delivered to both users.
With $\mathbb E[\mathbf s\mathbf s^H]=\mathbf I_2$, the array signal is $\mathbf x=\beta\mathbf W\mathbf H^{-1}\mathbf s$, and its total power is constrained to $P$. The received signal is interference-free up to the common scaling factor $\beta$, leading to the same SINR for both users.
Define the beam Gram matrix $\mathbf G=\mathbf W^H\mathbf W$. Normalizing the full precoder accounts for the generally non-orthogonal analog beams~\cite{ref_elaa_access}:

\begin{equation}
    \beta^2=\frac{P}{\operatorname{tr}(\mathbf H^{-H}\mathbf G\mathbf H^{-1})},\qquad \mathrm{SINR}_k=\frac{\beta^2}{\sigma_n^2}.
    \label{eq:zf_sinr}
\end{equation}
For cross-scheme conditioning, we use $\bar{\mathbf H}=\mathbf H\mathbf G^{-1/2}$. The ZF power cost is $\|\mathbf W\mathbf H^{-1}\|_F^2=\sum_i\sigma_i^{-2}(\bar{\mathbf H})$, which emphasizes weak singular modes. Both users have the same SINR; hence $R_{\rm sum}=2\log_2(1+\mathrm{SINR})$.
\subsubsection{Metric and Protocol}
Our primary metric is the SINR gain:
\begin{equation}
    \Delta\mathrm{SINR}_{\mathrm{dB}}=10\log_{10}\!\left(\frac{\mathrm{SINR}_P}{\mathrm{SINR}_A}\right),
    \label{eq:delta_sinr}
\end{equation}
where subscripts $P$ and $A$ denote the proposed Pearcey-inspired method and the baseline method, respectively.
We fix $P=1$ and $\sigma_n^2=10^{-3}$, giving a nominal total transmit-to-noise ratio of $30$~dB. The actual received SINR also depends on the effective channel.

\subsection{Fair Comparison Protocol}
\label{subsec:protocol}
Both designs follow the same comparison protocol:
\begin{itemize}
    \item \textbf{Baseline:} Uses free-space phase-conjugate focusing toward each user.
    \item \textbf{Pearcey (Proposed):} Designs beams ignoring the obstruction, using a calibrated quartic aperture phase.
    \item \textbf{Free-Space Calibration:} The quartic beam is calibrated with $M(x)=1$, without obstruction information.
    \item \textbf{Identical Evaluation:} Both methods use the same array, propagation model, obstruction, noise variance, total array power, and digital ZF rule.
\end{itemize}

\section{Pearcey-Inspired Wavefront Design and Calibration}
\label{sec:wavefront_design}

To mitigate performance degradation caused by obstructions without relying on explicit obstruction knowledge, we propose a wavefront shaping strategy inspired by the Pearcey catastrophe integral.
This section details the beam formulation, parameterization, and calibration procedure.

\subsection{Baseline: Conventional Obstruction-Unaware Focusing}
As a benchmark, we use phase-only spherical focusing weights designed independently for each user in free space.
At array element $x_n=(n-(N_\text{elem}+1)/2)d$, the weight is
\begin{equation}
    w_{k,n}^{(A)}=\frac{1}{\sqrt{N_\text{elem}}}\exp\!\left(j\Phi_{\text{focus}}(x_n;z_k)\right),
    \label{eq:baseline_A}
\end{equation}
with the exact spherical focusing phase $\Phi_{\text{focus}}(x; z_k) = \frac{2\pi}{\lambda} \sqrt{x^2 + z_k^2}$.
This baseline concentrates energy near the target in free space; partial blockage alters its effective multi-user response.

\subsection{Pearcey-Inspired Quartic Wavefront}
Pearcey diffraction motivates quartic phase shaping, with related beams studied in coherent and partially coherent optical fields~\cite{Ring2012Auto,Cao2025Diffraction}.
The canonical Pearcey integral has phase $t^4+u t^2+v t$, where $u$ and $v$ are control parameters~\cite{Ring2012Auto}.
Inspired by this, we retain the quartic component on top of conventional spherical focusing. The $x^4$ dependence concentrates the added phase variation toward the aperture edges while preserving symmetry about $x=0$, yielding a Pearcey-inspired wavepacket centered on the user axis.
The proposed array weight is
\begin{equation}
    w_{k,n}^{(P)}=w_{k,n}^{(A)}\exp\!\left[j\left(\delta a_2x_n^2+\alpha_4x_n^4\right)\right],
    \label{eq:pearcey_beam}
\end{equation}
where $\alpha_4x_n^4$ introduces the quartic phase structure, and $\delta a_2x_n^2$ is an auxiliary quadratic term used for free-space calibration (see Sec.~\ref{subsec:calibration}).

\begin{figure}[!t]
    \centering
    \subfloat[Free-space calibration]{\includegraphics[width=0.95\linewidth]{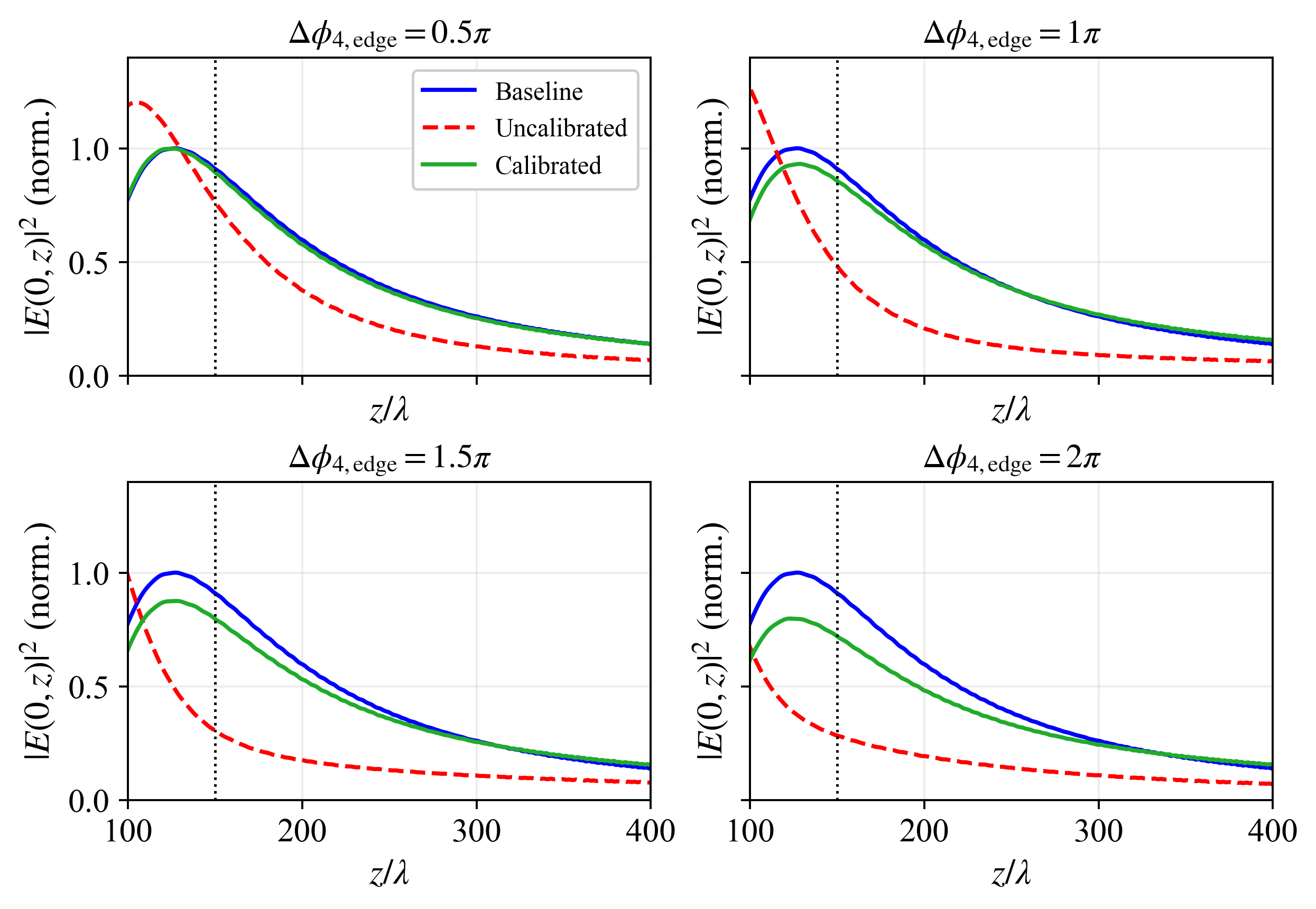}\label{fig:calibration_a}}\\[2pt]
    \subfloat[Sensitivity to $\Delta\phi_{4,\text{edge}}$]{\includegraphics[width=0.7\linewidth]{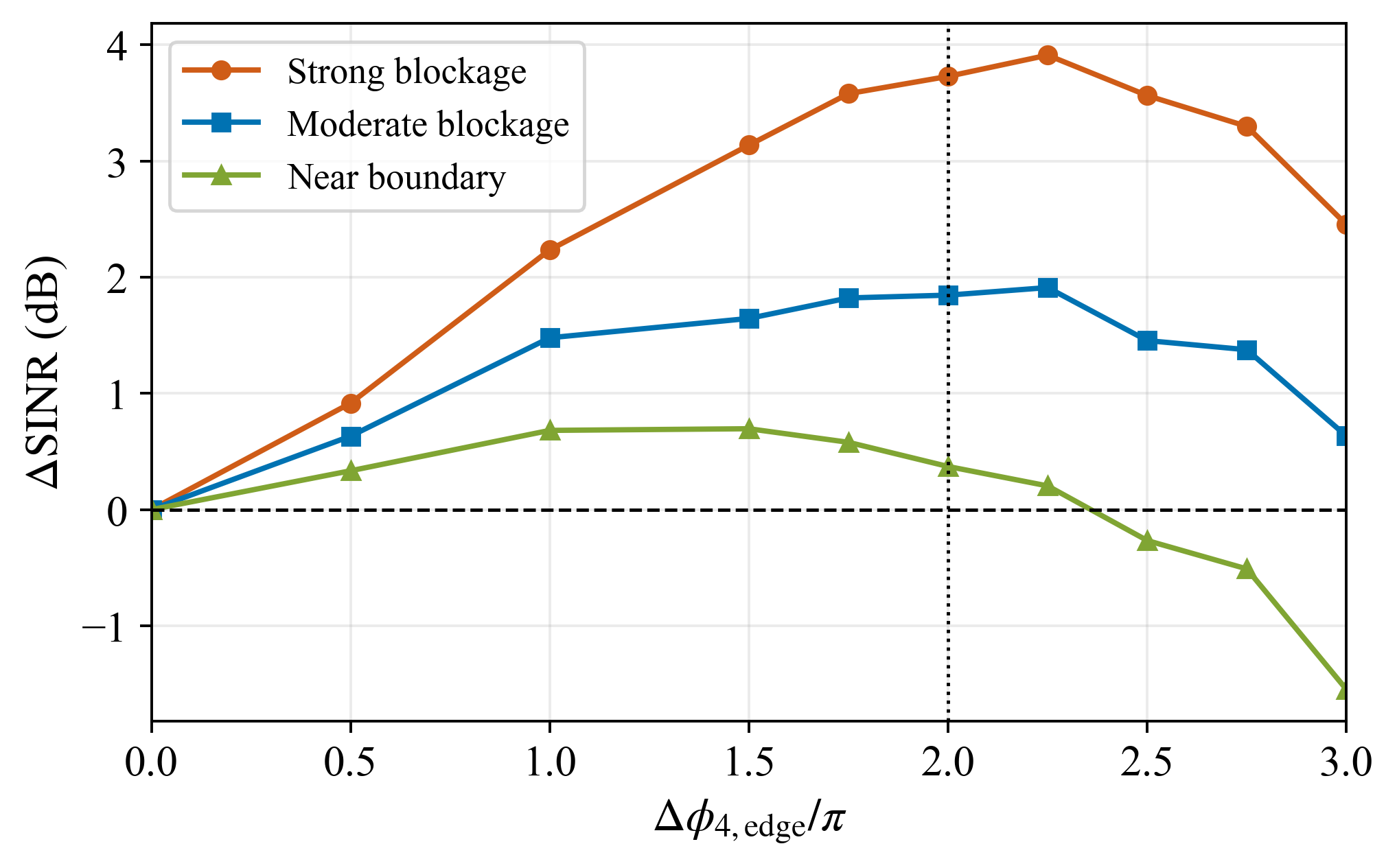}\label{fig:calibration_b}}
    \caption{\textbf{(a)} Free-space axial intensities for $\Delta\phi_{4,\text{edge}}\in\{0.5\pi,\pi,1.5\pi,2\pi\}$, normalized to the baseline peak. Calibration maximizes intensity at $z_1=150\lambda$ (dotted line). At $2\pi$, the target-intensity penalty is $1.02$~dB. \textbf{(b)} Quartic-strength sensitivity at $(\mu,W_{\text{obs}}/r_F)\approx(0.373,0.902)$, $(0.813,0.673)$, and $(2.060,0.379)$. The fixed $2\pi$ design is within $0.33$~dB of the best sampled strength at each point.}
    \label{fig:calibration}
\end{figure}

\subsection{Design Insight: Normalization and Parameterization}
The effect of the quartic coefficient $\alpha_4$ depends on the phase accumulated across the finite aperture. We therefore parameterize the quartic strength through its \textit{edge phase swing}, which directly expresses the imposed phase excursion.
For $x_{\text{edge}}=\pm D/2$, the edge phase swing is
\begin{equation}
    \Delta\phi_{4,\text{edge}} \triangleq \alpha_4 (D/2)^4 \quad \Rightarrow \quad \alpha_4 = \frac{\Delta\phi_{4,\text{edge}}}{(D/2)^4}.
    \label{eq:alpha4_def}
\end{equation}
\textbf{Design Choice:} We select $\Delta\phi_{4,\text{edge}}=2\pi$ based on the trade-off between on-axis efficiency and performance under central blockage.
This value corresponds to a full phase cycle at the aperture edge. The same edge phase sets a common quartic coefficient for the two beams, while their spherical focusing and quadratic calibration terms account for the different user ranges.
As shown in Fig.~\ref{fig:calibration}(a), $\Delta\phi_{4,\text{edge}}=2\pi$ incurs a free-space target-intensity penalty of approximately $1.02$~dB. Smaller values, such as $0.5\pi$, yield a negligible penalty but provide less SINR gain at the strongly blocked point in Fig.~\ref{fig:calibration}(b).
Fig.~\ref{fig:calibration}(b) further shows that the chosen $2\pi$ design lies within $0.33$~dB of the best sampled strength at three representative geometries spanning strong blockage, moderate blockage, and the advantage boundary. This supports using one common phase strength across the evaluated geometries.

\subsection{Free-Space Calibration and Gain-Robustness Trade-off}
\label{subsec:calibration}

\subsubsection{The ``Focus Shift'' Challenge}
Adding a quartic phase term $\alpha_4x^4$ shifts the axial intensity distribution of the finite aperture.
Without quadratic correction, the field at the nominal user location can decrease as the axial peak moves toward the aperture. We therefore adjust an auxiliary quadratic phase to recover intensity at the intended target.
This adjustment is made by \textbf{free-space calibration}.
The auxiliary quadratic parameter $\delta a_2$ in~\eqref{eq:pearcey_beam} is tuned to maximize the on-axis intensity at the user's nominal location $z_k$ under free-space propagation:
\begin{equation}
    \delta a_2^\star = \argmax_{\delta a_2} \left| E_{\mathbf{w}_k^{(P)}}(0, z_k) \big|_{M=1} \right|^2,
    \label{eq:calibration}
\end{equation}
where the field is evaluated with unit-norm array weights through the free-space propagator~\eqref{eq:fresnel_prop}, with $M(x)=1$. For each user, the search starts with 61 uniformly spaced candidates in $\delta a_2\in[-1.5,1.5]k_0/(2z_k)$. If the best candidate approaches an interval edge, the interval is expanded by a factor of 1.5, for at most three search rounds.
Each beam is calibrated independently in free space before evaluating the blocked multi-user channel.

\subsubsection{Gain vs.\ Robustness Trade-off}
Fig.~\ref{fig:calibration} visualizes the calibration results for different quartic strengths.
As $\Delta\phi_{4,\text{edge}}$ increases from $0.5\pi$ to $2\pi$, the calibrated intensity at the nominal target decreases from $-0.07$ to $-1.02$~dB relative to the baseline at that target. The calibrated axial maxima occur at approximately $123$--$128\lambda$, with the nominal user location fixed at $z_1=150\lambda$.
The phase-only weights preserve the array amplitude profile and total element power. The quartic term changes the relative phases of the aperture contributions and redistributes intensity in the propagated field.
Under a finite central obstruction, the surviving field comes from the two open portions of the screen. These contributions reach the user axis with phases determined by the incident wavefront and the remaining propagation distance. The calibrated quartic phase changes their superposition, allowing on-axis reception to be sustained beyond the obstruction.
For the two users at different ranges, this field reconstruction also changes the distinctness of their effective channel signatures. A stronger weakest channel mode reduces the power cost of separating the users with ZF, connecting the optical field redistribution to simultaneous communication performance.

\section{Numerical Results and Discussion}
\label{sec:results}

\begin{figure}[!t]
    \centering
    \includegraphics[width=\linewidth]{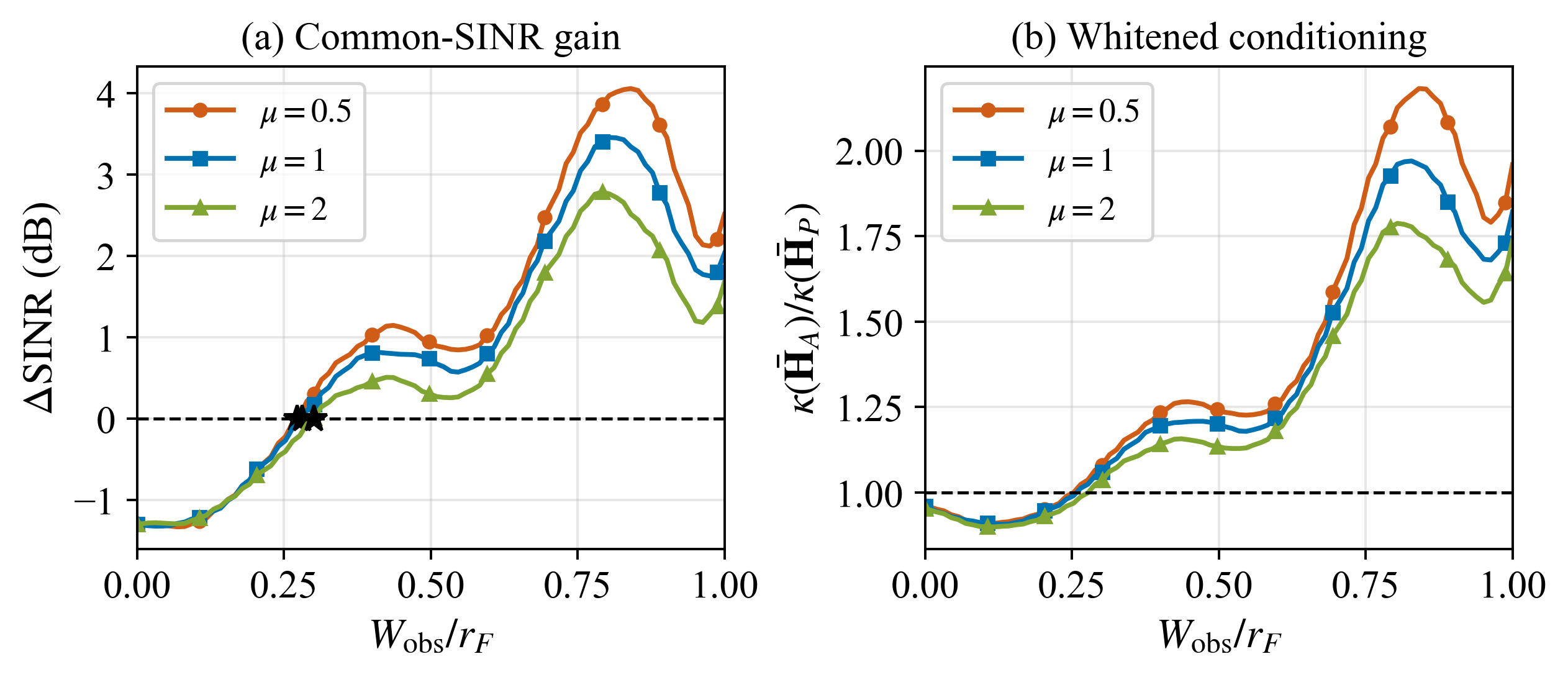}
    \caption{Performance comparison versus normalized obstruction width $W_{\text{obs}}/r_F$ for $\mu \in \{0.5, 1.0, 2.0\}$, at $\eta=0.5$.
    \textbf{(a)} Common-SINR gain over the baseline; stars mark zero crossings estimated by linear interpolation between samples.
    \textbf{(b)} Power-whitened condition-number ratio $\kappa(\bar{\mathbf H}_A)/\kappa(\bar{\mathbf H}_P)$. Values above one indicate improved conditioning.}
    \label{fig:1d_slices}
\end{figure}

\begin{figure}[!t]
    \centering
    \subfloat[Advantage map at $\eta=0.5$]{\includegraphics[width=0.48\linewidth]{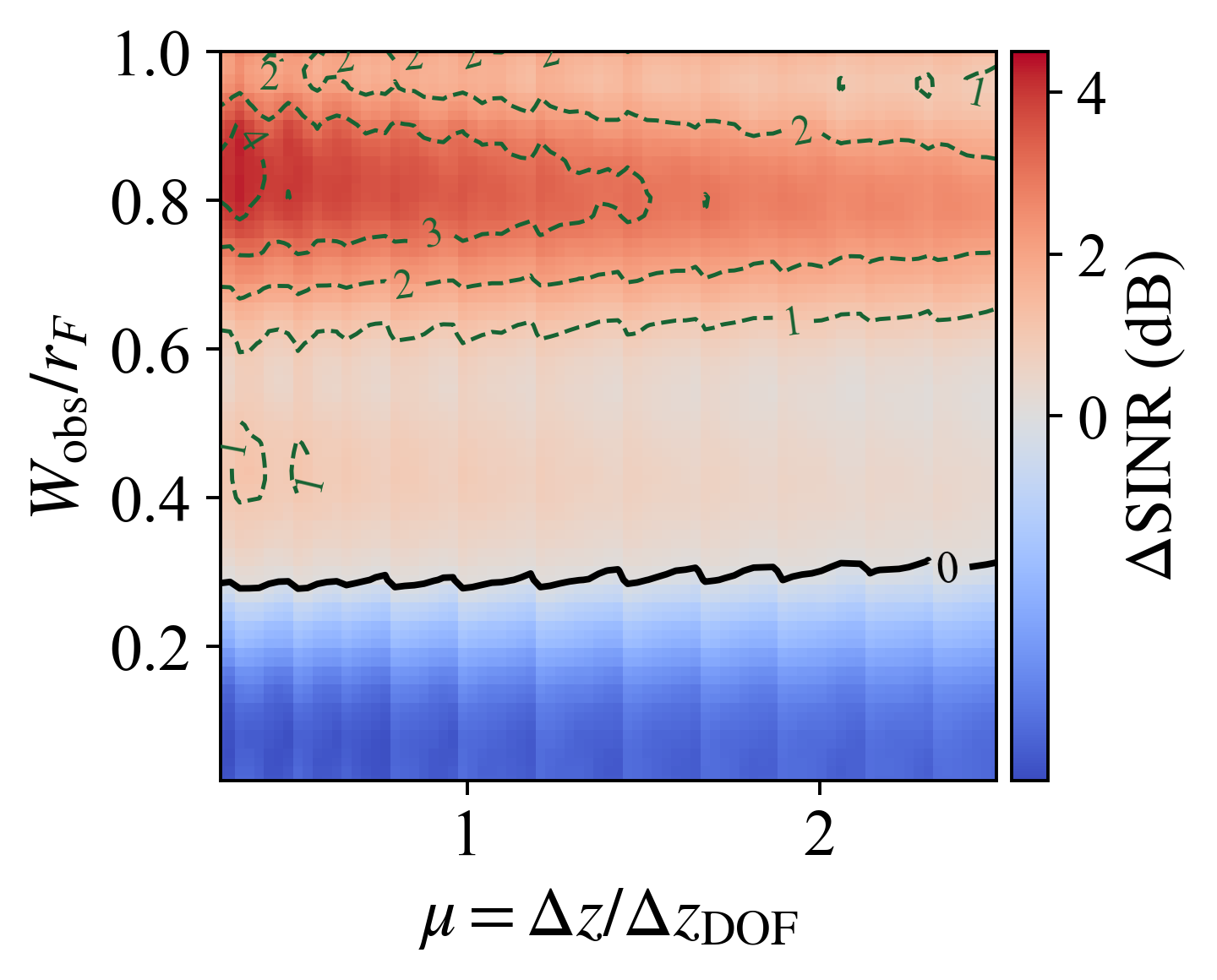}\label{fig:heatmap_a}}
    \hfil
    \subfloat[0\,dB contours per $\eta$]{\includegraphics[width=0.48\linewidth]{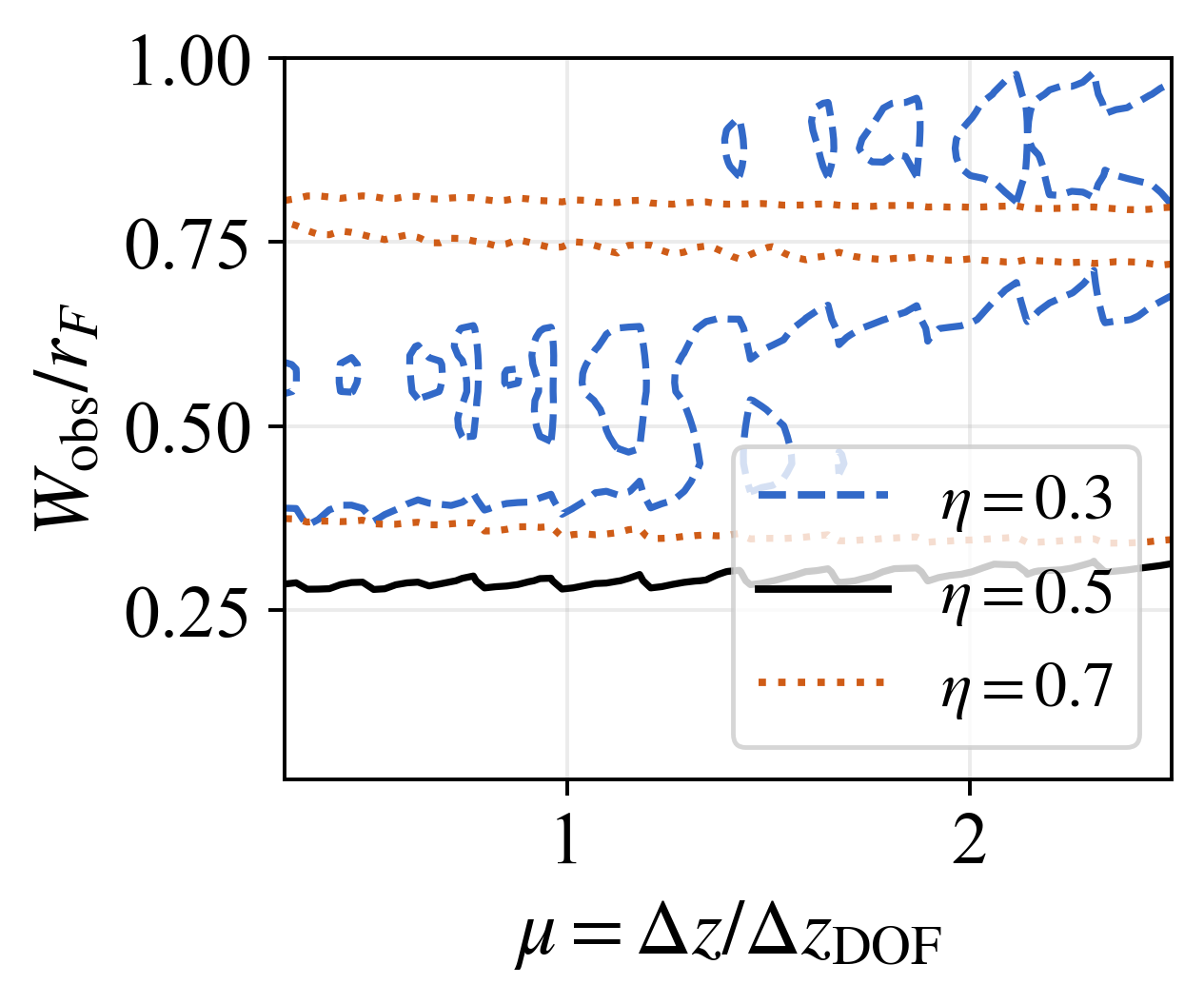}\label{fig:heatmap_b}}
    \caption{\textbf{(a)} Common-SINR advantage at $\eta=0.5$. The black contour marks zero gain; green dashed contours mark $1$, $2$, $3$, and $4$~dB. \textbf{(b)} Zero-gain contours computed separately for $\eta\in\{0.3,0.5,0.7\}$. Both panels use the raw sampled results without smoothing; the boundary varies with user separation and obstruction position.}
    \label{fig:heatmap}
\end{figure}

We evaluate the proposed Pearcey-inspired beamforming strategy using the fairness protocol defined in Sec.~\ref{subsec:protocol}.
The system parameters are $N_{\text{elem}}=64$, $d=0.49\lambda$, $z_1=150\lambda$, $\sigma_n^2=10^{-3}$, and $P=1$.
The 1-D slices in Fig.~\ref{fig:1d_slices} use $\mu\in\{0.5,1,2\}$ at the representative obstruction position $\eta=0.5$. The maps in Fig.~\ref{fig:heatmap} use an $81\times81$ grid over $\mu\in[0.3,2.5]$ and $W_{\text{obs}}/r_F\in[0.02,1]$ at each stated $\eta$. A calibration-resolution check at ten geometries with $\eta=0.5$ gives a maximum gain change of $0.075$~dB when the number of candidates increases from 61 to 961.

\subsection{Impact of Obstruction Size}
Fig.~\ref{fig:1d_slices} plots the SINR gain and the channel condition number ratio as a function of $W_{\text{obs}}/r_F$ for $\mu \in \{0.5, 1.0, 2.0\}$.

The curves exhibit an obstruction-width threshold for positive SINR gain.
In the unobstructed regime ($W_{\text{obs}}=0$), the proposed beam pair has a common-SINR penalty of approximately $1.30$~dB on all three slices. Conventional focusing benefits from concentrating the field near each target, while the quartic phase trades some focal intensity for a different spatial field distribution.

As obstruction size increases, the two wavefronts produce different coherent sums through the open regions of the screen. This changes both the received field strengths and the separability of the effective user channels.
The zero crossings estimated by linear interpolation are $W_{\text{obs}}/r_F\approx0.273$, $0.281$, and $0.300$ for $\mu=0.5$, $1$, and $2$, respectively. At $W_{\text{obs}}/r_F=1$, the corresponding gains are $2.51$, $2.03$, and $1.67$~dB.
The crossover occurs at a smaller obstruction width for more closely spaced users: their overlapping focal responses make ZF more sensitive to the weakest channel mode, increasing the benefit of quartic shaping.

Fig.~\ref{fig:1d_slices}(b) shows the corresponding improvement in power-whitened channel conditioning. The ratio $\kappa(\bar{\mathbf H}_A)/\kappa(\bar{\mathbf H}_P)$ broadly follows the SINR-gain trend, reaching approximately $1.96$, $1.82$, and $1.74$ at $W_{\text{obs}}/r_F=1$ for $\mu=0.5$, $1$, and $2$, respectively.
The largest ratio among these three slices occurs at the smallest $\mu$, where the baseline has less range separation available to distinguish the users. The quartic profile improves the relative strength of the weak mode, reducing the ZF inversion cost. The absolute singular-value scale also enters this cost through~\eqref{eq:zf_sinr}.

\subsection{Advantage Map}
Fig.~\ref{fig:heatmap}(a) presents the heatmap of the SINR gain over the $(\mu, W_{\text{obs}}/r_F)$ plane at the representative obstruction position $\eta=0.5$.
The strongest gains occur for central blockage with $W_{\text{obs}}/r_F\gtrsim0.75$ and closely spaced users with $\mu\lesssim0.6$. The peak sampled gain is $4.29$~dB at $(\mu,W_{\text{obs}}/r_F)=(0.355,0.8775)$.
For moderate obstruction, $W_{\text{obs}}/r_F\in[0.4,0.7]$, gains range from approximately $0.08$ to $2.48$~dB over the tested $\mu$ values at $\eta=0.5$.
For well-separated users, conventional focusing provides stronger range discrimination, reducing the relative benefit of quartic shaping. Positive-gain regions remain, with their locations determined by the diffracted field.
Fig.~\ref{fig:heatmap}(b) shows positive-gain regions for $\eta\in\{0.3,0.5,0.7\}$ with different boundaries. Moving the obstruction changes the incident field and the remaining propagation distances, altering the interference at the users and hence the advantage region.

The common-SINR gain can be interpreted through the ZF power cost.
Partial blockage can make the effective channel vectors of co-angular users highly correlated, increasing the ZF power cost $\sum_i\sigma_i^{-2}(\bar{\mathbf H})$. At fixed transmit power, this reduces the common scaling factor $\beta$ and the SINR delivered to both users.
At the peak-gain point, the quartic profile increases the whitened smallest singular value by a factor of $1.64$ and reduces the whitened condition number by a factor of $2.27$, preserving stronger, more distinct range signatures under blockage.
This improvement in effective-channel separability outweighs the focal-intensity penalty in the strong-blockage regime. The lower ZF power cost allows a higher common received signal level under the same transmit-power budget.

\section{Conclusion}
\label{sec:conclusion}

This letter proposed an obstruction-unaware wavefront shaping strategy using Pearcey-inspired quartic phase profiles to enhance multi-user RNF connectivity.
Free-space calibration adjusts the quadratic phase to maximize intensity at the nominal user position, while digital ZF evaluates simultaneous service under a common array-power budget.
Numerical results identify an advantage region in normalized obstruction width and user range separation, with a peak sampled ZF common-SINR gain of $4.29$~dB when strong central blockage coincides with closely spaced users.
Future work will investigate adaptive tuning of the quartic strength and the extension to two-dimensional apertures and three-dimensional propagation geometries.

\end{document}